\documentclass[12pt,cite,epsf,epsfig]{article}
\usepackage{epsfig}

\begin{document}

\author{S. Dev\thanks{dev5703@yahoo.com} $^{a, b}$, Radha Raman Gautam\thanks{gautamrrg@gmail.com} $^a$ and Lal Singh\thanks{lalsingh96@yahoo.com} $^a$}
\title{Broken $S_3$ Symmetry in the Neutrino Mass Matrix and Non-Zero $\theta_{13}$}
\date{$^a$\textit{Department of Physics, Himachal Pradesh University, Shimla 171005, India.}\\
\smallskip
$^b$\textit{Theoretical Physics Division, Institute of High Energy Physics, Chinese Academy of Sciences, P.O. Box 918, Beijing 100049, China.}}

\maketitle
\begin{abstract}
We study the effects of breaking $S_3$ symmetry in the neutrino mass matrix for the masses and mixing matrix of neutrinos. At zeroth order the model gives degenerate neutrino masses and accommodates tribimaximal mixing. We introduce perturbations in terms of a small and complex parameter. The perturbations are introduced in a manner such that the $S_3$ symmetry is broken by its elements in the same representation. Successive perturbations introduce mass splitting, sizable non-zero reactor mixing angle and CP violation. This scheme of breaking $S_3$ symmetry can reproduce a relatively large reactor mixing angle as suggested by the recent T2K results. The effective neutrino mass is predicted to be large which is testable in the ongoing and forthcoming neutrinoless double beta decay experiments.  
\end{abstract}

\section{Introduction}
After the discovery of neutrino oscillations, there has been considerable progress in determining the values of the neutrino mass-squared differences and the mixing angles which relate the neutrino mass eigenstates to the flavor eigenstates. Recently, T2K \cite{1}, MINOS \cite{2} and Double Chooz \cite{3} experiments have given hints of a relatively large 1-3 mixing angle. This has given a fresh impetus to the construction of neutrino mass models which can accommodate a non-zero value of $\theta_{13}$. Recently many papers have appeared which reproduce the relatively large value of the reactor mixing angle \cite{4}. The mixing pattern in the quark sector comprises of small mixing angles while in the lepton sector the mixing is described by two large angles and one small angle. Discrete symmetries have been extensively used in the past to obtain two large and one small mixing angle in the lepton sector. The most studied phenomenological Ans\"{a}tz for the neutrino mixing matrix arising from discrete symmetries was proposed by Harrison, Perkins and Scott \cite{5} known as tribimaximal mixing (TBM) is given by
\begin{equation}
U_{TBM} = \left(
\begin{array}{ccc}
\frac{-2}{\sqrt{6}} & \frac{1}{\sqrt{3}} & 0 \\
\frac{1}{\sqrt{6}}&
\frac{1}{\sqrt{3}}& \frac{-1}{\sqrt{2}}\\
\frac{1}{\sqrt{6}}&
\frac{1}{\sqrt{3}}& \frac{1}{\sqrt{2}}\\
\end{array}
\right).
\end{equation}
Non-Abelian flavor symmetries \cite{6} have been invoked to account for TBM. With the recent T2K results various neutrino mass models  based on TBM need to be suitably modified to accommodate a relatively large $\theta_{13}$. The smallest discrete non-Abelian group is $S_3$ which is the permutation group of three objects. A large number of papers \cite{7, 8} have presented detailed models based on $S_3$ symmetry. The permutation matrices in the three dimensional reducible representation are
\begin{equation}
S^{(1)} = \left(
\begin{array}{ccc}
1 &0& 0 \\
0&1&0\\
0&0&1\\
\end{array}
\right),
\end{equation} 
\begin{equation}
S^{(123)} = \left(
\begin{array}{ccc}
0 &0& 1 \\
1&0&0\\
0&1&0\\
\end{array}
\right),
S^{(132)}=\left(
\begin{array}{ccc}
0&1&0\\
0&0&1\\
1&0&0\\
\end{array}
\right),
\end{equation} 
\begin{equation}
S^{(12)}=\left(
\begin{array}{ccc}
0 &1& 0 \\
1&0&0\\
0&0&1\\
\end{array}
\right),
S^{(13)}=\left(
\begin{array}{ccc}
0 &0& 1 \\
0&1&0\\
1&0&0\\
\end{array}
\right),
S^{(23)} = \left(
\begin{array}{ccc}
1&0&0\\
0&0&1\\
0&1&0\\
\end{array}
\right),
\end{equation}
where the matrices in each equation belong to the same class of $S_3$. 
The invariance of neutrino mass matrix $M_\nu$ under $S_3$ requires
\begin{equation}
[S,M_\nu]=0
\end{equation}
where $S$ is any of the six permutation matrices given in Eqs. (2, 3, 4). The neutrino mass matrix invariant under $S_3$ is given by
\begin{equation}
M_\nu= aI+bD\\
\end{equation}
where
\begin{equation}
I=  \left(
\begin{array}{ccc}
1 &0& 0 \\
0&1&0\\
0&0&1\\
\end{array}
\right), 
D= \left(
\begin{array}{ccc}
1&1&1\\
1&1&1\\
1&1&1\\
\end{array}
\right)
\end{equation}
where $a$ and $b$ are, in general, complex and $D$ is called the democratic matrix. In this work we consider deviations from $S_3$ invariance of the neutrino mass matrix. We introduce the perturbations which break this symmetry in such a way that the perturbed neutrino mass matrix still satisfies the magic symmetry\cite{9,10}, for this we choose the perturbation matrices to be the $S_3$ group matrices. Indeed any linear combination of these matrices will preserve the magic symmetry. We consider the three different $Z_2$ subgroup matrices (Eq. 4) of $S_3$ as the perturbation matrices. Requiring symmetric mass matrices, the remaining $S_3$ matrices are not considered because $S^{(123)}$ and $S^{(132)}$ are not symmetric and $S^{(1)}$ does not change the $S_3$ invariance of the neutrino mass matrix. In Ref.\cite{9} the breaking of $S_3$ symmetry generated a non-zero $\theta_{13}$ but the deviation from zero was small. Moreover, the perturbation parameter was considered to be real for simplicity leading to the absence of Dirac-type CP violation. Here, we generalize that work by considering a complex perturbation  parameter which leads to Dirac-type CP violation in the model and, also, generates a relatively large $\theta_{13}$ consistent with the T2K results. We take the charged lepton mass matrix to be diagonal. If a horizontal symmetry exists it must, simultaneously, be a symmetry of the charged leptons as well as the neutrinos before the gauge symmetry breaking. After symmetry breaking when the fermions acquire mass, the charged lepton and the neutrino mass matrices should be constrained by different subgroups of the symmetry group in order to have non-zero mixing. We consider $S_3$ to be the residual group in the neutrino sector. For the charged lepton sector, $Z_3$ symmetry with the representation $diag(1,\omega,\omega^2)$ where $\omega = e^{i 2 \pi/3}$, can be taken as the residual symmetry which yields non degenerate diagonal charged lepton mass matrix \cite{11}. Before the spontaneous symmetry breaking both the left-handed charged leptons and neutrinos share a common doublet so they must be governed by the same horizontal group which in this case comes out to be $\Delta(54)$ \cite{12}. One may easily find that the gauge interactions of leptons remain invariant under the full group  $\Delta(54)$.
\section{$S_3$ invariant neutrino mass matrix}
The neutrino mass matrix (Eq.~6) can be diagonalised by any unitary matrix with a single trimaximal eigenvector, for example the TBM matrix $U_{TBM}$. This $S_3$ invariant neutrino mass matrix has eigenvalues $a$, $a+3b$ and $a$ respectively. This is contrary to the experimental data since the two degenerate eigenvalues correspond to the mass eigenvalues $m_1$ and $m_3$ while experimentally $m_1$ and $m_2$ have smaller mass difference. A possible solution to this problem was provided by Jora \textit{et al.}\cite{8} by introducing a $CP$ violating phase $\alpha$ between the complex vectors $a$ and $a+3b$. The phase $\alpha$ is adjusted to ensure equal magnitude of $a$ and $a+3b$, leading to degenerate mass spectrum for $S_3$ invariant $M_\nu$ at the zeroth order. The complex plane is oriented in a manner so that the parameter $b$ is completely imaginary, (Fig.(1)). The complex vector $a$ lies in the fourth quadrant and is given by
\begin{equation}
a=|a|e^{-i \alpha /2}.
\end{equation}
\begin{figure}
\begin{center}
\epsfig{file=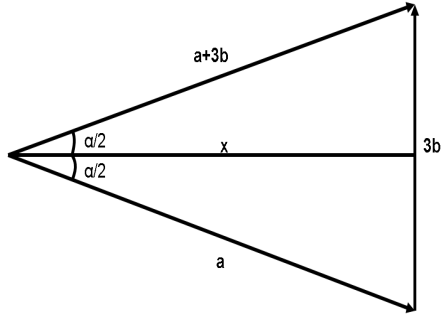,height=6.0cm,width=6.0cm}
\end{center}
\caption{Angle $\alpha$ between two equal length vectors $a$ and $a+3b$.}
\end{figure} 
The complex parameters $a$ and $b$ in terms of the real free parameter $x$ are
\begin{eqnarray}
|a|=x \sec (\alpha/2),\nonumber \\
|b|=\frac{2}{3}x \tan (\alpha/2).
\end{eqnarray}
Democratic charged lepton mass matrices of the form $bD$ and diagonal neutrino mass matrices of the form $aI$ have been discussed earlier \cite{13} in the context of $S_3$ symmetry. The possibility $a=0$ has, also, been discussed in the literature \cite{14}. $S_3$ flavor symmetry in the context of Type (I+II) see saw has been discussed in Ref.\cite{15}. A generic feature of this approach is a democratic charged lepton mass matrix and a diagonal neutrino mass matrix at the zeroth order. Democratic mass matrix in the neutrino sector with terms breaking $S_3$ symmetry to $\mu-\tau$ symmetry has been discussed in \cite{16}. In contrast, we consider the most general $S_3$ invariant neutrino mass matrices at the zeroth order in the flavor basis where the charged lepton mass matrix is diagonal.
\section{Deviations from $S_3$ symmetry}
In this section, we introduce explicit perturbations to the $S_3$ invariant neutrino mass matrix $M_\nu$ and study their effects on neutrino masses and mixings. We introduce the perturbations in terms of a small, complex and dimensionless parameter $\lambda e^{i\psi}$. The perturbed neutrino mass matrix upto second order perturbation is given by
\begin{equation}
M'_\nu = M_\nu +\mu (\lambda e^{i\psi} [S^p]+\lambda^2 e^{2i\psi} [S^q])
\end{equation}
where $M_\nu$ is invariant under $S_3$ and  $S^p$ and $S^q$ can be any of $S^{(12)}$, $S^{(13)}$ and $S^{(23)}$ such that $p \neq q$ and $\mu$ is the real parameter with dimensions of mass and magnitude of the order of one in the units of the absolute mass scale. The matrices $S^{(12)}$, $S^{(13)}$ and $S^{(23)}$ given in Eq. (4) belong to the same class of $S_3$. In this scheme of perturbation, the $S_3$ symmetry is broken by its elements in the same representation. Even after introducing the second order perturbation, the neutrino mass matrix holds the magic property which leads to a trimaximal eigenvector for the perturbed neutrino mass matrix $M'_\nu$. In the present case, the first order perturbation is taken to be $\lambda e^{i\psi} S^{(23)}$ \cite{9}. This leads to the following neutrino mass matrix after the first order perturbation with $\lambda e^{i\psi}S^{(23)}$:
\begin{equation}
M^{(1)}_\nu = \left(
\begin{array}{ccc}
a+b+ \mu \lambda e^{i\psi} & b & b \\
b & a+b & b+ \mu  \lambda e^{i\psi} \\
b & b+ \mu \lambda e^{i\psi}& a+b \\
\end{array}
\right).
\end{equation}
This mass matrix (Eq.~11) now leads uniquely to TBM mixing but now the degeneracy in the neutrino masses is broken.
The complex neutrino masses are given by
\begin{center}
\begin{eqnarray}
\begin{array}{ccc}
m_1= (x+\mu \lambda \cos \psi)+ i(\mu \lambda \sin \psi - x \tan(\frac{\alpha}{2})) ,\\
m_2=(x+\mu \lambda \cos \psi)+ i(\mu \lambda \sin \psi + x \tan(\frac{\alpha}{2})) ,\\
m_3=(x+\mu \lambda \cos \psi)- i(\mu \lambda \sin \psi + x \tan(\frac{\alpha}{2})) .\\
\end{array}
\end{eqnarray}
\end{center}

To the first order in $\lambda$, the neutrino masses are given by
\begin{center} 
\begin{eqnarray}
\begin{array}{ccc}
|m_1|=x \sec \frac{\alpha}{2}+\mu \lambda \cos(\frac{\alpha}{2}+\psi) ,\\
|m_2|=x\sec \frac{\alpha}{2}+\mu \lambda \cos(\frac{\alpha}{2}-\psi),\\
|m_3|=x\sec \frac{\alpha}{2}-\mu \lambda \cos(\frac{\alpha}{2}+\psi).\\
\end{array}
\end{eqnarray}
\end{center} 
The second order perturbation leads to deviations from TBM and generates a non-zero $\theta_{13}$. This can be done either by adding $\lambda^2 e^{2i\psi} S^{(12)}$ or $\lambda^2 e^{2i\psi} S^{(13)}$ as the second order perturbation. The two choices only differ in their predictions for $\theta_{23}$ and $\psi$. Here, we take $\lambda^2 e^{2i\psi} S^{(12)}$ as the second order perturbation. The perturbed neutrino mass matrix after second order perturbation becomes
\begin{equation}
M'_\nu = \left(
\begin{array}{ccc}
a+b+ \mu \lambda e^{i\psi} & b+\mu \lambda^2 e^{2i\psi} & b \\
b+\mu \lambda^2 e^{2i\psi} & a+b& b+\mu \lambda e^{i\psi} \\
b & b+\mu \lambda e^{i\psi} & a+b+\mu \lambda^2 e^{2i\psi} \\
\end{array}
\right).
\end{equation}
It is clear that $M'$ still satisfies the magic property and thus has a trimaximal eigenvector $(\frac{1}{\sqrt{3}},\frac{1}{\sqrt{3}},\frac{1}{\sqrt{3}})^T$ \cite{10}. The corresponding mixing matrix can be obtained by the application of a general 1-3 rotation to the TBM matrix from the right 
$U_{TBM} U_{13}$ where 
\begin{equation}
U_{13} = \left(
\begin{array}{ccc}
\sqrt{1-v^2-w^2} & 0 & v+iw \\
0 & 1 & 0 \\
-v+iw & 0 & \sqrt{1-v^2-w^2} \\
\end{array}
\right).
\end{equation}
\begin{figure}
\begin{center}
\epsfig{file=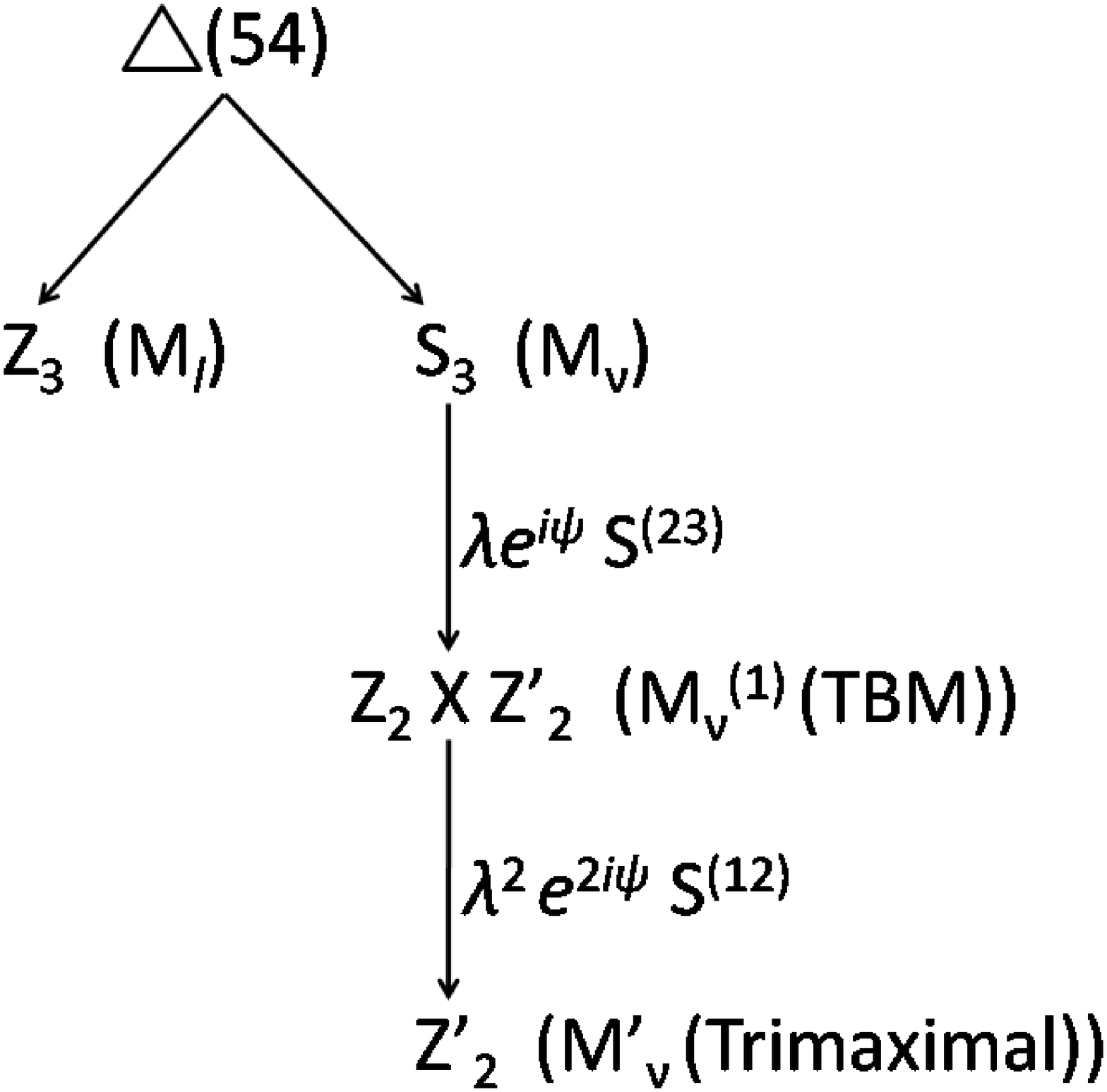,height=6.0cm,width=7.0cm}
\end{center}
\caption{Symmetry breaking chain.}
\end{figure}  
Now we discuss the symmetry breaking chain, (Fig.(2)). After spontaneous symmetry breaking, the full group $\Delta(54)$ breaks to $Z_3$ in the charged lepton sector and $S_3$ in the neutrino sector. Note that the $S_3$ invariant neutrino mass matrix $M_\nu$ possesses accidental magic symmetry ($Z'_2$) \textit{i.e.} $M_\nu$ is also invariant under $Z'_2$
\begin{equation}
Z'_2=\frac{1}{3}\left(
\begin{array}{ccc}
1 & -2 & -2 \\
-2 & 1 & -2 \\
-2 & -2 & 1 \\
\end{array}
\right).
\end{equation}
 The first order perturbation breaks $S_3$ to the Klein's four-group $Z_2 \times Z'_2$, where $Z_2$ is the $\mu-\tau$ exchange symmetry. Although  Klein's four-group is not a subgroup of $S_3$ but the $S_3$ invariant $M_\nu$ has additional $Z'_2$ symmetry which allows for this breaking pattern. $Z_2 \times Z'_2$ is the symmetry group for TBM in the perturbed neutrino mass matrix $M_\nu^{(1)}$. The second order perturbation breaks $Z_2 \times Z'_2$ to $Z'_2$ and, thus, the perturbed neutrino mass matrix $M'_\nu$ has a trimaximal eigenvector.\\
Diagonalizing the perturbed neutrino mass matrix $M'_\nu$ using $U_{TBM} U_{13}$, the neutrino masses to the second order in $\lambda$ are given by
\begin{equation}
\begin{array}{ccc}
|m_1| \approx  x \sec \frac{\alpha}{2}[ 1+\cos ^2 (\frac{\alpha}{2}) \lbrace \frac{\lambda \mu \cos(\frac{\alpha}{2}+\psi)\sec(\frac{\alpha}{2})}{x}+\frac{\lambda ^2 (\mu^2 - \mu^2 \cos ^2 (\frac{\alpha}{2}+\psi) -x \mu \cos(\frac{\alpha}{2}+\psi)\sec(\frac{\alpha}{2})}{2x^2}\rbrace ],\\
|m_2| \approx x \sec \frac{\alpha}{2}[ 1+\cos ^2 (\frac{\alpha}{2}) \lbrace \frac{\lambda \mu \cos(\frac{\alpha}{2}-\psi)\sec(\frac{\alpha}{2})}{x}+\frac{\lambda ^2 (\mu^2 - \mu^2 \cos ^2 (\frac{\alpha}{2}-\psi) +2 x \mu \cos(\frac{\alpha}{2}-2 \psi)\sec(\frac{\alpha}{2})}{2x^2}\rbrace ],\\
|m_3| \approx x \sec \frac{\alpha}{2}[ 1+\cos ^2 (\frac{\alpha}{2}) \lbrace \frac{-\lambda \mu \cos(\frac{\alpha}{2}+\psi)\sec(\frac{\alpha}{2})}{x}+\frac{\lambda ^2 (\mu^2 - \mu^2 \cos ^2 (\frac{\alpha}{2}+\psi) -x \mu \cos(\frac{\alpha}{2}+\psi)\sec(\frac{\alpha}{2})}{2x^2}\rbrace ].
\end{array}
\end{equation}
The neutrino masses are made real positive by the phase matrix
\begin{small}
\begin{equation}
P=\left(
\begin{array}{ccc}
e^{-i \tau} & 0 & 0 \\
0 & e^{-i \sigma} & 0 \\
0 & 0 & e^{-i \rho} \\
\end{array}
\right)
\end{equation}
\end{small}
where
\begin{small}
\begin{equation}
\begin{large}
\begin{array}{ccc}
\tau \approx \frac{1}{2}\tan^{-1}(\frac{- x\tan (\frac{\alpha}{2})+\lambda \mu \sin(\psi)-\frac{\lambda^2}{2} \mu \sin(2 \psi)}{x+ \lambda \mu \cos(\psi)-\frac{\lambda^2}{2} \mu \cos(2 \psi)}),\\
\sigma \approx \frac{1}{2}\tan^{-1}(\frac{x\tan (\frac{\alpha}{2})+\lambda \mu \sin(\psi)+\lambda^2 \sin(2 \psi)}{x+ \lambda \mu \cos(\psi)+\lambda^2 \mu \cos(2 \psi)}),\\
\rho \approx \frac{1}{2}\tan^{-1}(\frac{- x\tan (\frac{\alpha}{2})-\lambda \mu \sin(\psi)+\frac{\lambda^2}{2}\mu \sin(2 \psi)}{x- \lambda \mu \cos(\psi)+\frac{\lambda^2}{2} \mu \cos(2 \psi)}).
\end{array}
\end{large}
\end{equation}
\end{small}
Thus, the full mixing matrix is given by
\begin{equation}
U'=U_{TBM} U_{13} P.
\end{equation}
The solar and atmospheric mass-squared differences to second order in $\lambda$ are given by
\begin{equation}
\begin{array}{cc}
\triangle m_{21}^2 \equiv (m_2^2-m_1^2 ) \approx x\lambda \mu (3 \lambda \cos(2 \psi)+2(2+\lambda \cos(\psi))\sin(\psi)\tan(\frac{\alpha}{2})) ,\\
\triangle m_{23}^2 \equiv (m_2^2-m_3^2) \approx x\lambda \mu (\lambda \cos(2 \psi)+\cos(\psi)(4+6 \lambda \sin(\psi)\tan(\frac{\alpha}{2}))).
\end{array}
\end{equation}
In the earlier analysis \cite{9}, the mass-squared differences were independent of the phase $\alpha$ implying that it is a Majorana-type phase as lepton number conserving neutrino oscillations do not depend on this phase. This was the consequence of taking the perturbation parameter $\lambda$ to be real. In the present study, however, both the phases $\alpha$ and $\psi$ appear in the above mass-squared differences and the phase $\alpha$ can no longer be identified as a pure Majorana phase. If we take the phase $\psi$ to be zero, $\alpha$ decouples from the mass-squared differences and its Majorana character is restored. The neutrino mixing matrix after the second order perturbation is given by
\begin{equation}
U' = \left(
\begin{array}{ccc}
-\sqrt{\frac{2}{3}} \left(1-\frac{v^2}{2}\right) & \frac{1}{\sqrt{3}} &
-\sqrt{\frac{2}{3}} (-v-i w) \\
\frac{1-\frac{v^2}{2}}{\sqrt{6}}-\frac{v-i w}{\sqrt{2}} & \frac{1}{\sqrt{3}} &
\frac{-v-i w}{\sqrt{6}}-\frac{1-\frac{v^2}{2}}{\sqrt{2}} \\
\frac{1-\frac{v^2}{2}}{\sqrt{6}}+\frac{v-i w}{\sqrt{2}} & \frac{1}{\sqrt{3}} &
\frac{1-\frac{v^2}{2}}{\sqrt{2}}+\frac{-v-i w}{\sqrt{6}}
\end{array}
\right)
\end{equation}
where
\begin{eqnarray}
v=\frac{\sqrt{3} \lambda \cos(\frac{\alpha}{2}+2 \psi) \sec(\frac{\alpha}{2} +\psi)}{4},\\
w=-\frac{\sqrt{3}\lambda^2 \mu \cos(\frac{\alpha}{2})\sec(\frac{\alpha}{2}+\psi) \sin(\psi)}{4 x}.
\end{eqnarray}
The neutrino mixing angles upto the second order perturbation are given by
\begin{equation}
\sin^2\theta_{13} \approx \frac{\lambda ^2 \cos(\frac{\alpha}{2}+2 \psi)^2 \sec(\frac{\alpha}{2} +\psi)^2}{8 },
\end{equation}
\begin{equation}
\sin^2\theta_{12} \approx \frac{1}{3}+\frac{\lambda ^2 \cos(\frac{\alpha}{2}+2 \psi)^2 \sec(\frac{\alpha}{2} +\psi)^2}{24 },
\end{equation}
and
\begin{equation}
\sin^2\theta_{23} \approx \frac{1}{2}+\frac{\lambda \cos(\frac{\alpha}{2}+2 \psi) \sec(\frac{\alpha}{2} +\psi)}{4 }.
\end{equation}
The CP violation in neutrino oscillation experiments can be described through a rephasing invariant quantity, $J_{CP}$ \cite{17} with $J_{CP}=Im(U'_{e1} U'_{\mu2} U'^*_{e2} U'^*_{\mu1})$. The expression for $J_{CP}$ to second order in $\lambda$ is given by
\begin{equation}
J_{CP} \approx -\frac{\lambda^2 \mu \cos(\frac{\alpha}{2})\sec(\frac{\alpha}{2}+\psi) \sin(\psi)}{12 x}.
\end{equation}
Since $J_{CP}$ contains two different phases neither of them can be directly identified with the Dirac-type phase $\delta$ in the standard PDG representation \cite{18}, rather some combination of these phases corresponds to $\delta$.
As pointed out earlier we have two choices for second order perturbation matrices. The above expressions were obtained for $\lambda^2 e^{2i\psi} S^{(12)}$ as the second order perturbation. If we use $\lambda^2 e^{2i\psi} S^{(13)}$ as the second order perturbation, the predictions for all the parameters remain the same except for $\theta_{23}$ and $J_{CP}$. The atmospheric mixing angle is now given by
\begin{equation}
\sin^2\theta_{23} \approx \frac{1}{2}-\frac{\lambda \cos(\frac{\alpha}{2}+2 \psi) \sec(\frac{\alpha}{2} +\psi)}{4 }
\end{equation}
while $J_{CP}$ has the same value as above but with an opposite sign. The atmospheric mixing angle has the same deviation from maximal mixing as in the earlier case but in the opposite direction.
\section{Numerical results}
The experimental constraints on neutrino parameters at 1, 2 and 3$\sigma$ \cite{19} are given in Table 1.
\begin{table}
\begin{center}
\begin{tabular}{|c|c|}
\hline Parameter & mean$^{(+1 \sigma, +2 \sigma, +3 \sigma)}_{(-1 \sigma, -2 \sigma, -3 \sigma)}$ \\
\hline $\Delta m_{21}^{2} [10^{-5}eV^{2}]$ & $7.59_{(-0.18,-0.35,-0.50)}^{(+0.20,+0.40,+0.60)}$ \\ 
\hline $\Delta m_{31}^{2} [10^{-3}eV^{2}]$ & $2.50_{(-0.16,-0.25,-0.36)}^{(+0.09,+0.18,+0.26)}$, \\&
$(-2.40_{(-0.09,-0.17,-0.27)}^{(+0.08,+0.18,+0.27)})$ \\ 
\hline $\sin^2 \theta_{12}$ & $0.312_{(-0.015,-0.032,-0.042)}^{(+0.017,+0.038,+0.048)}$ \\ 
\hline $\sin^2 \theta_{23}$ & $0.52_{(-0.07,-0.11,-0.13)}^{(+0.06,+0.09,+0.12)}$, \\& $(0.52_{(-0.06,-0.10,-0.13)}^{(+0.06,+0.09,+0.12)})$ \\ 
\hline $\sin^2 \theta_{13}$ & $0.013_{(-0.005,-0.009,-0.012)}^{(+0.007,+0.015,+0.022)}$,\\& $(0.016_{(-0.006,-0.011,-0.015)}^{(+0.008,+0.015,+0.019)})$ \\ 
\hline 
\end{tabular}
\caption{Current Neutrino oscillation parameters from global fits \cite{19}. The upper (lower) row corresponds to Normal (Inverted) Spectrum, with $\Delta m^2_{31} > 0$ ($\Delta m^2_{31} < 0$).}
\end{center}
\end{table}
The effective Majorana mass of the electron neutrino $M_{ee}$ which determines the rate of neutrinoless double beta (NDB) decay is given by \cite{20}
\begin{equation}
M_{ee}=|m_1U_{11}'^2+m_2U_{12}'^2+m_3U_{13}'^2  |.
\end{equation}
Part of the Heidelberg-Moscow collaboration claimed a signal in NDB decay corresponding to $M_{ee} = (0.11 - 0.56) eV$ at 95$\%$ C. L. \cite{21}. This claim was subsequently criticized in \cite{22}. The results reported in \cite{21} will be checked in the currently running and forthcoming NDB experiments. 
There are a large number of projects such as CUORICINO\cite{23}, 
CUORE \cite{24}, GERDA \cite{25}, MAJORANA \cite{26}, SuperNEMO \cite{27}, EXO \cite{28}, GENIUS \cite{29} which aim to achieve a sensitivity upto 0.01eV for $M_{ee}$.
Cosmological observations put an upper bound on the sum of neutrino masses
\begin{equation}
\Sigma = \sum_{i=1}^3 m_i.
\end{equation}
The WMAP data \cite{30} limit $\Sigma$ to be less than $0.67 eV$ at $95 \%$ C.L..
\begin{figure}
\begin{center}
\epsfig{file=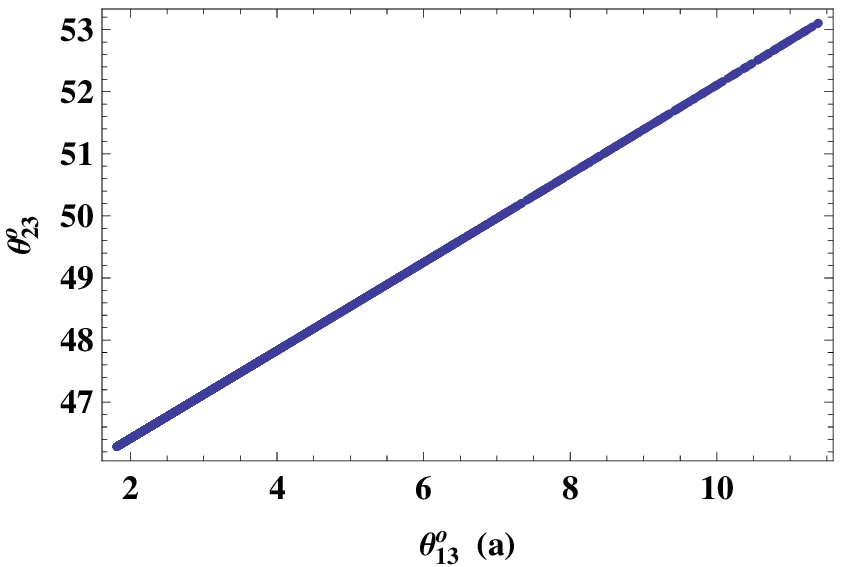,height=5.0cm,width=5.0cm}
\epsfig{file=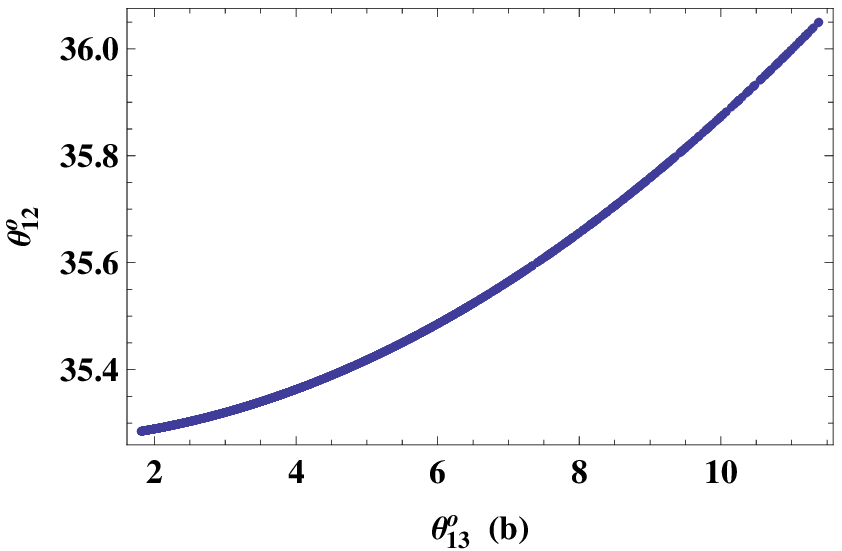,height=5.0cm,width=5.0cm}
\epsfig{file=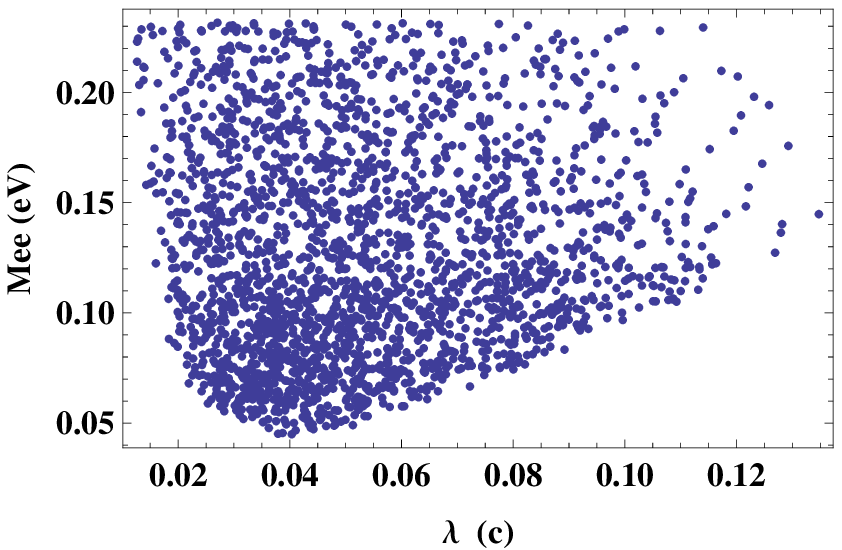,height=5.0cm,width=5.0cm}
\epsfig{file=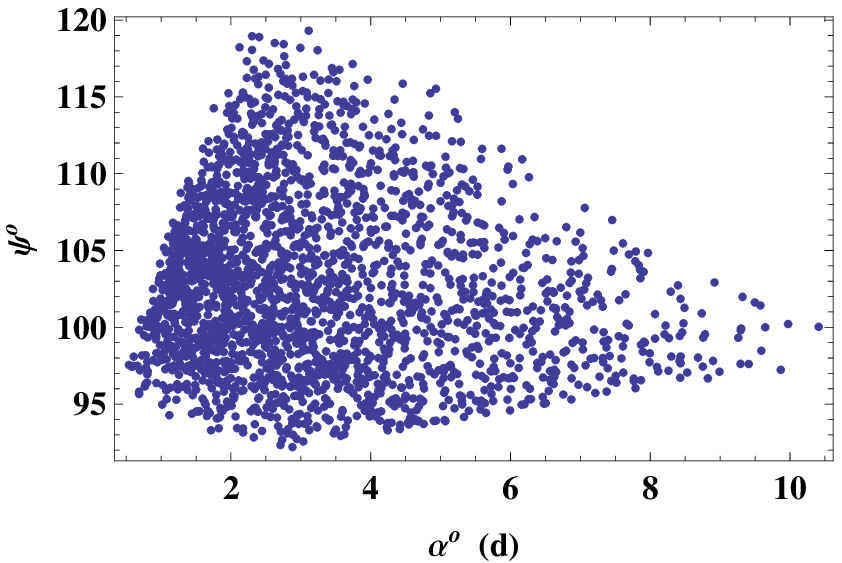,height=5.0cm,width=5.0cm}
\end{center}
\caption{Correlation plots for the Normal Spectrum.}
\end{figure}
For numerical analysis, we use the $3 \sigma$ ranges of neutrino oscillation parameters.  
For Normal Spectrum (NS), the parameters $\lambda$ and $\mu$ are constrained to lie in the range $(0.012-0.135)$ and $(0.1-2) eV$, $x$ is constrained to the range $(0.034-0.234)eV$ and the two phases $\alpha$ and $\psi$ have ranges $(0.5^o-10.5^o)$ and $(92^o-120^o)$ respectively. The effective Majorana mass $M_{ee}$ has the range $(0.043-0.232) eV$ while the CP violating parameter $J_{CP}$ lies in the range $(0-0.07)$.
For Inverted Spectrum (IS), the allowed ranges for the parameters  $\lambda$ and $\mu$ are $(0.011-0.115)$ and $(0.1-2) eV$ while $x$ can vary in the range $(0.032-0.233) eV$ and the phases $\alpha$ and $\psi$ lie in the range $(1.4^o-9.35^o)$ and $(57^o-86^o)$ respectively. $M_{ee}$ lies between $(0.065-0.235) eV$ and $J_{CP}$ lies between $((-0.07)-0)$. The deviations of the mixing angles for NS and IS are shown in Figs.(3a, 3b) and Figs.(4a, 4b) respectively.
\begin{figure}
\begin{center}
\epsfig{file=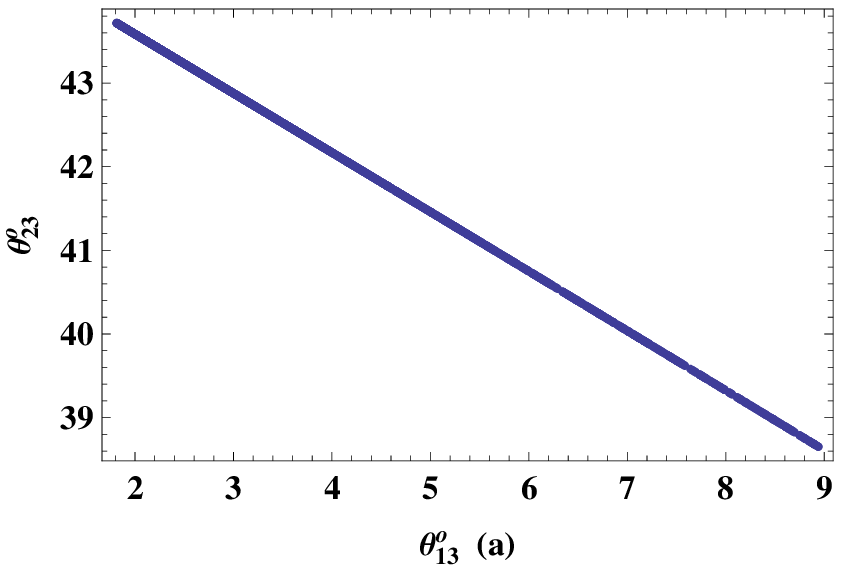,height=5.0cm,width=5.0cm}
\epsfig{file=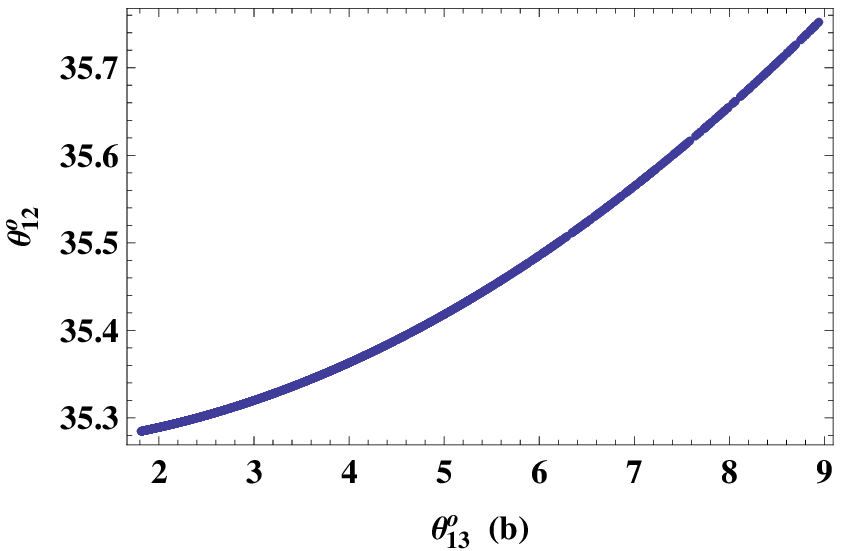,height=5.0cm,width=5.0cm}
\epsfig{file=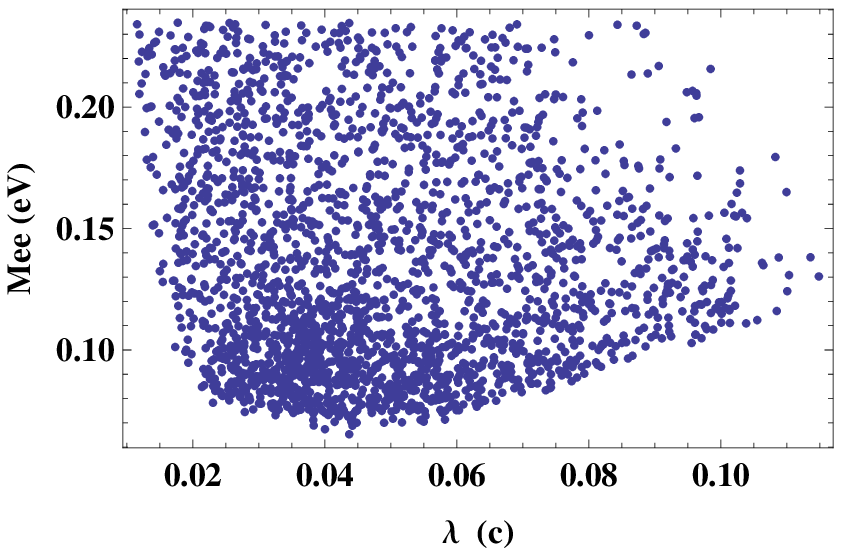,height=5.0cm,width=5.0cm}
\epsfig{file=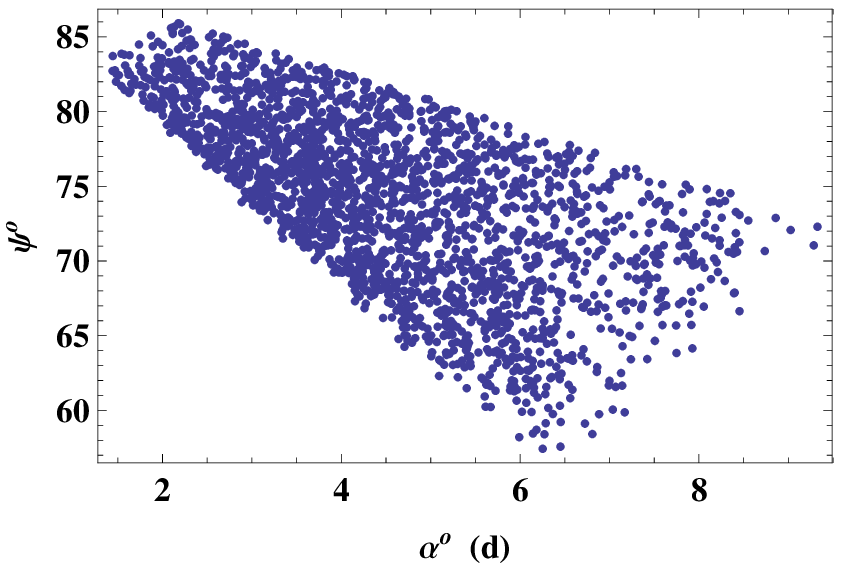,height=5.0cm,width=5.0cm}
\end{center}
\caption{Correlation plots for the Inverted Spectrum.}
\end{figure}
\section{Summary}
In the present work, we generalized the approach used in Ref.\cite{9} for breaking $S_3$ symmetry in the neutrino mass matrix by considering a small but complex perturbation parameter. Our starting point is a zeroth order mass matrix arranged to give three degenerate neutrino masses. This mass matrix can be diagonalized by e.g. the TBM mixing matrix. Mass splittings are introduced by adding perturbations to the zeroth order neutrino mass matrix. This also leads to a relatively large reactor mixing angle and CP violation in the neutrino mass matrix. The perturbation matrices are chosen to be the $S_3$ group matrices in the three dimensional reducible representation belonging to the same class. Replacing the real perturbation parameter used in the earlier work \cite{9} by a complex perturbation parameter has profound consequences for the phenomenology of neutrino masses and mixings. A real perturbation parameter produced very small reactor mixing angle and there was no CP violation in the model. To accommodate the recent T2K results indicating a relatively large reactor mixing angle it becomes necessary to consider a complex perturbation parameter which also generates CP violation. The neutrino masses are found to be quasidegenerate and a large effective neutrino mass is predicted which is testable in the ongoing and forthcoming experiments for neutrinoless double beta decay.

\textbf{\textit{\Large{Acknowledgements}}}

S. D. gratefully acknowledges the warm hospitality extended by the Theoretical Physics Division,
Institute of High Energy Physics, Beijing. The research work of S. D. and L. S. is supported by the University Grants Commission, Government of India \textit{vide} Grant No. 34-32/2008
(SR). R. R. G. acknowledges the financial support provided by the Council for Scientific and Industrial Research (CSIR), Government of India.


\begin{thebibliography}{99}
\bibitem{1} K. Abe et al. [T2K collaboration], \textit{Phys. Rev. Lett.} \textbf{107}, 041801 (2011), arXiv:1106.2822 [hep-ex].
\bibitem{2} P. Adamson et al. [MINOS collaboration], \textit{Phys. Rev. Lett.} \textbf{107}, 181802 (2011), arXiv:1108.0015 [hep-ex].
\bibitem{3} Talk given by H. de Kerret at the Sixth International Workshop on Low Energy Neutrino Physics (LowNu11) at Seoul, Korea during November 9-12 (2011). 
\bibitem{4} Y. H. Ahn, Hai-Yang Cheng, S. Oh, \textit{Phys. Rev.} \textbf{D 83}, 076012 (2011), arXiv:1102.0879 [hep-ph]; Hong-Jian He, Fu-Rong Yin, \textit{Phys. Rev.} \textbf{D 84}, 033009 (2011), arXiv:1104.2654 [hep-ph]; Y. Shimizu, M. Tanimoto, A. Watanabe, \textit{Prog. Theor. Phys.} \textbf{126}, 81 (2011), arXiv:1105.2929 [hep-ph]; Zhi-zhong Xing,  arXiv:1106.3244 [hep-ph]; E. Ma, D. Wegman, \textit{Phys. Rev. Lett.} \textbf{107}, 061803 (2011), arXiv:1106.4269 [hep-ph]; Ya-juan Zheng, Bo-Qiang Ma, arXiv:1106.4040 [hep-ph]; S. Zhou, \textit{Phys. Lett.} \textbf{B 704}, 291 (2011), arXiv:1106.4808 [hep-ph]; T. Araki, \textit{Phys. Rev.} \textbf{D 84}, 037301 (2011), arXiv:1106.5211 [hep-ph]; N. Haba, R. Takahashi, \textit{Phys. Lett.} \textbf{B 702}, 388 (2011), arXiv:1106.5926 [hep-ph]; S. Morisi, K. M. Patel, E. Peinado, arXiv:1107.0696 [hep-ph]; D. Meloni \textit{JHEP} \textbf{1110}, 10 (2011), arXiv:1107.0221 [hep-ph]; W. Chao, Ya-juan Zheng arXiv:1107.0738 [hep-ph]; H. Zhang, S. Zhou, arXiv:1107.1097 [hep-ph]; S. Dev, S. Gupta, R. R. Gautam, \textit{Phys. Lett.} \textbf{B 704}, 527 (2011), arXiv:1107.1125 [hep-ph]; X. Chu, M. Dhen, T. Hambye, arXiv:1107.1589 [hep-ph]; P. S. B. Dev, R. N. Mohapatra, M. Severson, arXiv:1107.2378 [hep-ph]; R. de A. Toorop, F. Feruglio, C. Hagedron, \textit{Phys. Lett.} \textbf{B 703}, 447 (2011), arXiv:1107.3486 [hep-ph]; S. Antusch, V. Maurer, arXiv:1107.3728 [hep-ph]; W. Rodejohann, H. Zhang, S. Zhou, arXiv:1107.3970 [hep-ph]; Y. H. Ahn, H. Y. Cheng, S. Oh, \textit{Phys. Rev.} \textbf{D 84}, 113007 (2011), arXiv:1107.4549 [hep-ph]; S. F. King, C. Luhn, \textit{JHEP } \textbf{1109}, 042 (2011), arXiv:1107.5332 [hep-ph]; D. Marzocca, S. T. Petcov, A. Romanino, M. Spinarth, arXiv:1108.0614 [hep-ph]; S. F. Ge, D. A. Dicus, W. W. Repko, arXiv:1108.0964 [hep-ph]; S. Kumar, \textit{Phys. Rev.} \textbf{D 84}, 077301 (2011), arXiv:1108.2137 [hep-ph]; F. Bazzocchi, arXiv:1108.2497 [hep-ph]; T. Araki, C, Q, Geng, arXiv:1108.3175 [hep-ph]; S. Antusch, S. F. King, C. Luhn, M. Spinarth, arXiv:1108.4278 [hep-ph]; H. Fritzsch, Zhi-zhong Xing, S. Zhou, \textit{JHEP } \textbf{1109}, 083 (2011), arXiv:1108.4534 [hep-ph];  A. Rashed, A. Datta, arXiv:1109.2320 [hep-ph]; P. O. Ludl, S. Morisi, E. Peinado, arXiv:1109.3393 [hep-ph]; S. Verma, \textit{Nucl. Phys.} \textbf{B 854}, 340 (2012) arXiv:1109.4228 [hep-ph]; D. Meloni, arXiv:1110.5210 [hep-ph]; S. Dev, S. Gupta, R. R. Gautam, L. Singh, \textit{Phys. Lett.} \textbf{B 706}, 168 (2011), arXiv:1111.1300 [hep-ph]; K. N. Deepthi, S. Gollu, R. Mohanta, arXiv:1111.2781 [hep-ph]; A. Rashed, arXiv:1111.3072 [hep-ph]; I. de Medeiros Varzielas, arXiv:1111.3952 [hep-ph]; S. F. King, C. Luhn, arXiv:1112.1959 [hep-ph]; T. Araki, Y. F. Li, arXiv:1112.5819 [hep-ph];  S. Gupta, A. S. Joshipura, K. M. Patel, arXiv:1112.6113 [hep-ph]; Gui-Jun Ding, arXiv:1201.3279 [hep-ph]; H. Ishimori, T. Kobayashi, arXiv:1201.3429 [hep-ph]. 
\bibitem{5} P. F. Harrison, D. H. Perkins and W. G. Scott, \textit{Phys. Lett.} \textbf{B 530}, 167 (2002), hep-ph/0202074; Zhi-zhong Xing, \textit{Phys. Lett.} \textbf{B 533}, 85 (2002), hep-ph/0204049; P. F. Harrison and W. G. Scott, \textit{Phys. Lett.} \textbf{B 535}, 163 (2002), hep-ph/0203209.
\bibitem{6} E Ma and G Rajasekaran, \textit{Phys. Rev.} \textbf{D 64}, 113012 (2001), hep-ph/0106291; G. Altarelli and F. Feruglio, \textit{Nucl. Phys.} \textbf{B 720}, 64 (2005), hep-ph/0504165; \textit{Nucl. Phys.} \textbf{B 741}, 215 (2006), hep-ph/0512103; C. S. Lam, \textit{Phys. Rev. Lett.} \textbf{101}, 121602 (2008), arXiv:0804.2622 [hep-ph]; For recent reviews and list of references, see e.g. G. Altarelli and F. Feruglio, \textit{Rev. Mod. Phys.} \textbf{82}, 2701 (2010), arXiv:1002.0211 [hep-ph]; H. Ishimori, T. Kobayashi, H. Ohki, H. Okada, Y. Shimizu and M. Tanimoto, \textit{Prog. Theor. Phys. Suppl.} \textbf{183}, 1-163 (2010), arXiv:1003.3552 [hep-ph].
\bibitem{7} L. Wolfenstein, \textit{Phys. Rev.} \textbf{D 18}, 958 (1978); S. Pakvasa and H. Sugawara, \textit{Phys. Lett.} \textbf{B 73}, 61 (1978); \textbf{82}, 105 (1979); E. Derman, \textit{Phys. Rev.} \textbf{D 19}, 317 (1979); E. Durman and H. S. Tsao, \textit{Phys. Rev.} \textbf{D 20}, 1207 (1979); M. Fukugita, M. Tanimoto and T. Yanagida, \textit{Phys. Rev.} \textbf{D 57}, 4429 (1998), hep-ph/9709388; H. Fritzsch and Z. Z. Xing, \textit{Phys. Rev.} \textbf{D 61}, 073016 (2000), hep-ph/9909304; E. Ma, \textit{Phys. Rev.} \textbf{D 61}, 033012 (2000), hep-ph/9909249; M. Tanimoto, \textit{Phys. Lett.} \textbf{B 483}, 417 (2000), hep-ph/0001306; E. Ma and G. Rajasekaran, \textit{Phys. Rev.} \textbf{D 64}, 113012 (2001), hep-ph/0106291; P. F. Harrison and W. G. Scott, \textit{Phys. Lett.} \textbf{B 557}, 76 (2003), hep-ph/0302025; S. L. Chen, M. Frigerio and E. Ma, \textit{Phys. Rev.} \textbf{D 70}, 073008 (2004), [Erratum-\textit{ibid.} 70, 079905 (2004)], hep-ph/0404084; F. Caravaglios and S. Morisi, arXiv: hep-ph/0503234; W. Grimus and L. Lavoura, \textit{JHEP} \textbf{0508}, 013 (2005), hep-ph/0504153; R. N. Mohapatra, S. Nasri and H. B. Yu, \textit{Phys. Lett.} \textbf{B 639}, 318 (2006), hep-ph/0605020; N. Haba, K. Yoshioka, \textit{Nucl. Phys.} \textbf{B 739}, 254 (2006), hep-ph/0511108; M. Picariello, \textit{Int. J. Mod. Phys.} \textbf{A 23}, 4435 (2008), hep-ph/0611189; Y. Koide, \textit{Eur. Phys. J.} \textbf{C 50}, 809 (2007), hep-ph/0612058; A. Mondragon, M. Mondragon and E. Peinado, \textit{Phys. Rev.} \textbf{D 76}, 076003 (2007), arXiv:0706.0354 [hep-ph]; C. Y. Chen, L. Wolfenstein, \textit{Phys. Rev.} \textbf{D 77}, 093009 (2008), arXiv:0709.3767 [hep-ph]; F. Feruglio and Y. Lin, \textit{Nucl. Phys.} \textbf{B 800}, 77 (2008), arXiv:0712.1528 [hep-ph]; M. Mitra, S. Choubey, \textit{Phys. Rev.} \textbf{D 78}, 115014 (2008), arXiv:0806.3254 [hep-ph]; Duane A. Dicus, Shao-Feng Ge, Wayne W Repko , \textit{Phys. Rev.} \textbf{D 82}, 033005 (2010), arXiv:1004.3266 [hep-ph]; Zhi-zhong Xing, Deshan Yang, Shun Zhou, \textit{Phys. Lett.} \textbf{B 690}, 304 (2010), arXiv:1004.4234 [hep-ph]; D. Meloni, S. Morisi, E. Peinado, \textit{J. Phys.} \textbf{G, 38}, 015003 (2011), arXiv:1005.3482 [hep-ph]; S. Zhou, \textit{Phys. Lett.} \textbf{B 704}, 291 (2011), arXiv:1106.4808 [hep-ph].
\bibitem{8} R. Jora, S. Nasri and J. Schechter, \textit{Int. J. Mod. Phys.} \textbf{A 21}, 5875 (2006), hep-ph/0605069; R. Jora, J. Schechter and M. Naeem Shahid, \textit{Phys. Rev.} \textbf{D 80}, 093007 (2009), [Erratum-\textit{ibid.} \textbf{82}, 079902 (2010)], arXiv:0909.4414 [hep-ph]; Renata Jora, Joseph Schechter and M. Naeem Shahid, \textit{Phys. Rev.} \textbf{D 82}, 053006 (2010) arXiv:1006.3307 [hep-ph].
\bibitem{9}S. Dev, S. Gupta, R. R. Gautam, \textit{Phys. Lett.} \textbf{B 702}, 28-33 (2011), arXiv:1106.3873 [hep-ph].
\bibitem{10} P. F. Harrison and W. G. Scott, \textit{Phys. Lett.} \textbf{B 557}, 76 (2003), hep-ph/0302025; C. S. Lam, \textit{Phys. Lett.} \textbf{B 640}, 260 (2006), hep-ph/0606220; W. Grimus, L. Lavoura, \textit{JHEP}, \textbf{0809}, 106 (2008), arXiv:0809.0226 [hep-ph]; C. H. Albright, W. Rodejohann \textit{Eur. Phys. J.} \textbf{C 62}, 599 (2009), arXiv:0812.0436 [hep-ph]; S. Kumar, \textit{Phys. Rev.} \textbf{D 82}, 013010 (2010), arXiv:1007.0808[hep-ph].
\bibitem{11} C. S. Lam, \textit{Phys. Rev.} \textbf{D 78}, 073015 (2008), arXiv:0809.1185 [hep-ph]; 
\bibitem{12} J. A. Escobar and C. Luhn, \textit{J. Math. Phys.} \textbf{50}, 013524 (2009), arXiv:0809.0639[hep-th]; H. Ishimori, T. Kobayashi, H. Okada, Y. Shimizu and M. Tanimoto, \textit{JHEP}, \textbf{0904}, 011 (2009), arXiv:0811.4683 [hep-ph]; \textit{ibid.} \textbf{0912}, 054 (2009), arXiv:0907.2006 [hep-ph].
\bibitem{13} H. Fritzsch, Z. Z. Xing, \textit{Phys. Lett.} \textbf{B 440}, 313 (1998), hep-ph/9808272; H. Fritzsch, Z. Z. Xing \textit{Phys. Lett.} \textbf{B 598}, 237 (2004), hep-ph/0406206
\bibitem{14} H. Fritzsch, Z. Z. Xing, \textit{Phys. Lett.} \textbf{B 372}, 265 (1996), hep-ph/9509389.
\bibitem{15} W. Rodejohann, Z. Z. Xing, \textit{Phys. Lett.} \textbf{B 601}, 176 (2004), hep-ph/0408195.
\bibitem{16} J. S. D\'{i}az, V. A. Kosteleck\'{y}, \textit{Phys. Lett.} \textbf{B 700}, 25 (2011), arXiv:1012.5985 [hep-ph].
\bibitem{17} C. Jarlskog, \textit{Phys. Rev. Lett.} \textbf{55}, 1039 (1985).
\bibitem{18} Particle Data Group (2010), K. Nakamura, \textit{et. al.}, \textit{J. Phys.} \textbf{G}: \textit{Nucl. Part. Phys.} \textbf{37}, 075021. 
\bibitem{19} T. Schwetz, M. Tortola, J. W. F. Valle, \textit{New J. Phys.}, \textbf{13}, 109401 (2011), arXiv:1108.1376 [hep-ph].
\bibitem{20} For recent reviews see e.g. F. T. Avignone III, S. R. Elliott, J. Engel, \textit{Rev. Mod. Phys.} \textbf{80}, 481 (2008), arXiv:0708.1033 [nucl-ex]; S. M. Bilenky, \textit{Phys. Part. Nucl.} \textbf{41}, 690 (2010), arXiv:1001.1946  [hep-ph]; W. Rodejohann, \textit{Int. J. Mod. Phys.} \textbf{E, 20}, 1833 (2011), arXiv:1106.1334 [hep-ph]; J. J. Gomez-Cadenas, J. Martin-Albo, M. Mezzetto, F. Monrabal, M. Sorel, arXiv:1109.5515  [hep-ex].
\bibitem{21} H. V. Klapdor- Kleingrothaus, A. Dietz, H. L. Harney, I. V. Krivosheina, \textit{Mod. Phys. Lett.} \textbf{A 16}, 2409 (2001), hep-ph/0202018.
\bibitem{22}F. Feruglio, A. Strumia, F. Vissani, \textit{Nucl. Phys.} \textbf{B 637}, 345 (2002), hep-ph/0201291; C. E. Aalseth \textit{et. al.}, \textit{Mod. Phys. Lett.} \textbf{A 17}, 1475 (2002), hep-ph/0202018.
\bibitem{23} C. Arnaboldi \textit{et al.}, (CUORICINO collaboration), \textit{Phys. Lett. B} \textbf{584}, 260 (2004).
\bibitem{24} C. Arnaboldi \textit{et al.}, \textit{Nucl. Instrum. Methods Phys. Res. Sect.} \textbf{A 518}, 775 (2004).
\bibitem{25} I. Abt \textit{et al.}, (GERDA collaboration) hep-ex/0404039.
\bibitem{26} R. Gaitskell \textit{et. al.} [Majorana Collaboration] hep-ex/0311013.
\bibitem{27} A. S. Barabash [NEMO Collaboration], \textit{Czech. J. Phys.}, \textbf{52}, 567 (2002), nucl-ex/0203001.
\bibitem{28} M. Danilov \textit{et. al.}, \textit{Phys. Lett.} \textbf{B 480}, 12 (2000), hep-ex/0002003.
\bibitem{29} H. V. Klapdor- Kleingrothaus, \textit{et. al.}, \textit{Eur. Phys. J.} \textbf{A 12}, 147 (2001), hep-ph/0103062.
\bibitem{30} WMAP collaboration (E. Komatsu \textit{et al.}), \textit{Astrophys. J. Suppl.}, \textbf{180}, 330 (2009), arXiv:0803.0547 [astro-ph].


\end{thebibliography}
\end{document}